\documentclass{IEEEcsmag}

\usepackage[colorlinks,urlcolor=blue,linkcolor=blue,citecolor=blue]{hyperref}

\usepackage{upmath}

\usepackage{amsthm}
\usepackage{amsfonts}
\usepackage{amssymb}

\usepackage{cite}

  \usepackage[]{graphicx}
  \graphicspath{{images/}}
  \DeclareGraphicsExtensions{.pdf,.jpeg,.png}

\usepackage[cmex10]{amsmath}
\usepackage{algorithmic}
\usepackage{array}
\usepackage{multirow}

\usepackage{eqparbox}

\usepackage[]{caption}
\usepackage[font=footnotesize]{subfig}

\usepackage{stfloats}

\begin{document}


\title{FNR: A Similarity and Transformer-Based Approach to Detect Multi-Modal Fake News in Social Media}

\author{Faeze Ghorbanpour}
\affil{Department of Computer Engineering, Sharif University of Technology}

\author{Maryam Ramezani}
\affil{Department of Computer Engineering, Sharif University of Technology}

\author{MohammadAmin Fazli}
\affil{Department of Computer Engineering, Sharif University of Technology}

\author{Hamid R. Rabiee}
\affil{Department of Computer Engineering, Sharif University of Technology}

\markboth{Fake News Revealer}{Fake News Revealer}

\begin{abstract}
The availability and interactive nature of social media have made them the primary source of news around the globe. The popularity of social media tempts criminals to pursue their immoral intentions by producing and disseminating fake news using seductive text and misleading images. Therefore, verifying social media news and spotting fakes is crucial. This work aims to analyze multi-modal features from texts and images in social media for detecting fake news. We propose a Fake News Revealer (FNR) method that utilizes transform learning to extract contextual and semantic features and contrastive loss to determine the similarity between image and text. We applied FNR on two real social media datasets. The results show the proposed method achieves higher accuracies in detecting fake news compared to the previous works.
\end{abstract}

\maketitle

\chapterinitial{The introduction} The proliferation of social media has made it the first source of news for many people. News is published on social networks every day, and people, willingly or unwillingly, spread it from one channel to another, from person to person, and react to it. According to the study in \cite{news1}, the source of most Americans' news is through the Internet rather than newspapers or radio, and three-quarters of people said they receive news via email or social media. Additionally, 37 percent of people share news on Facebook and Twitter, suggesting that these two media have created a new way of sharing news. Social media news differs from other news sources like news agencies or microblogs. Usually, social media news is written by ordinary people, using informal language, being brief, and having poor-quality images. Our focus is on the news published on social media.

The two characteristics of fake news are: intentionally written and provable to be false, which distinguishes it from other kinds of news like rumors, satires and spams \cite{fake1}. Ease of use, sharing, and disseminating news by the public in social media can lead criminals to produce and publish fake news with malicious intent. These types of fraud can be perpetrated to mislead the public, harm an institution, person, government, or harm public and private stock markets. Meanwhile, the lack of awareness of the falsity of the news causes ordinary people to publish without knowledge. The scope of the news becomes wider and causes more significant damage, leading to distrust and disregard for the correct news and warnings work. It makes it difficult for reporters and journalists to cover correct and important news.

\begin{figure*}[!t]
\centerline{
	\subfloat[Real news]{
		\includegraphics[width=1.4in]{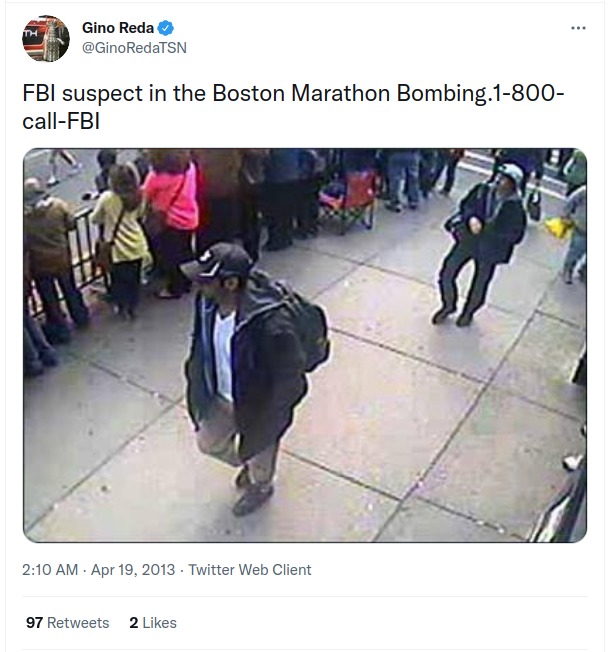}%
		\label{real1}
	}
\hfil
	\subfloat[Real news]{
		\includegraphics[width=1.4in]{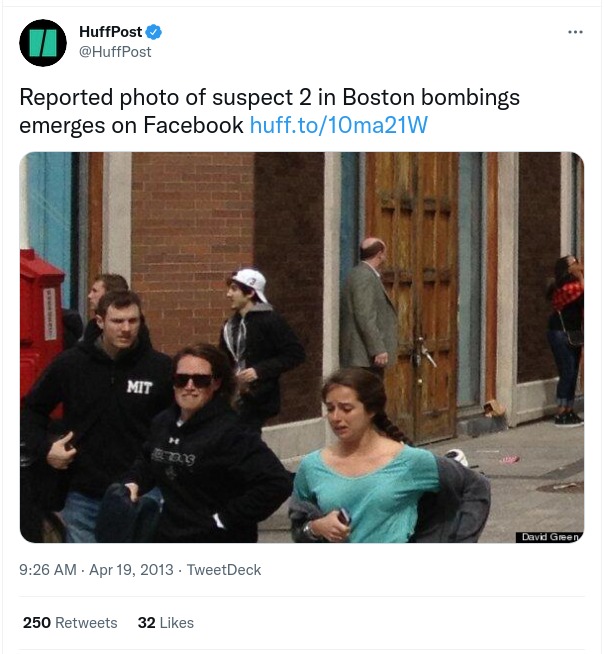}%
		\label{real2}
		}
\hfil
	\subfloat[Fake news]{
		\includegraphics[width=1.4in]{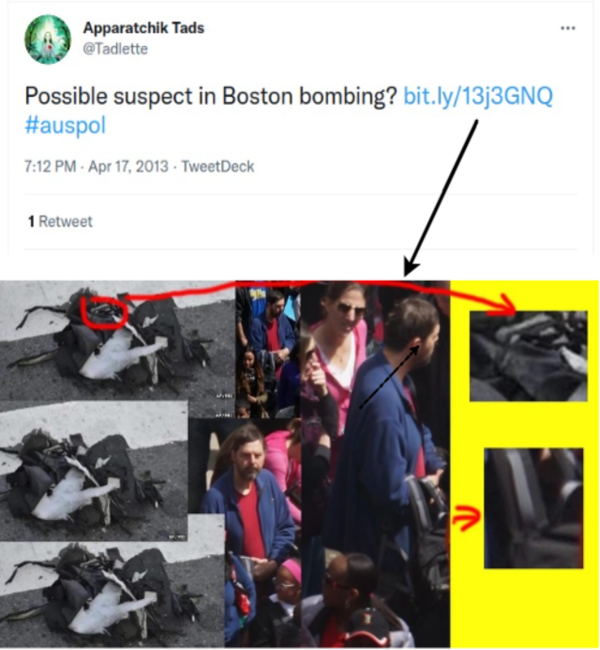}%
		\label{fake1}
		}
\hfil
	\subfloat[Fake news]{
		\includegraphics[width=1.4in]{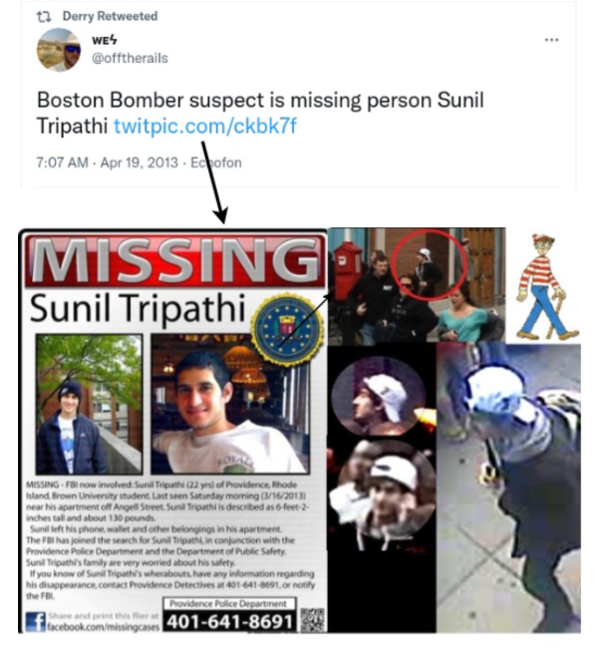}%
		\label{fake2}
		}
	}
\caption{Twitter posts about April 15, 2013 Boston terrorist attack}
\label{tweets}
\end{figure*}

Visual misinformation tends to involve much more straightforward deceptions. It is prevalent for old photos and videos to be repurposed as evidence of recent events. Using out-of-context images is another way to validate the news and gain trust. This means that they use the photo of the correct news in another event or context for fake news. Images play a pivotal role in influencing public opinion and creating false perceptions. According to psychological research \cite{image1}, when individuals see an image alongside a trivial statement such as turtles are deaf, they are more likely to believe it. In a simulated social media environment, a post that incorporates photos get more likes and shares, as well as people's perception that it is factual \cite{image2}. Hence, considering the visual features of news and the similarity relations between its image and text is essential for recognizing the truth of the news.

In social media, we do not always know the source of the news, but we can find the reaction of users against the news post. On Twitter, for example, people retweet or comment on tweets with or without intent. We only see retweets and comments, so determining their reliability is critical. These reactions can create fake news and divert attention from the main news, which prevents people from paying attention to the real story. For example, Figure \ref{tweets} shows some tweets related to a terrorist attack in Boston in 2013. As you can see, fake tweets are used to mislead readers, divert news, or further their agendas by exploiting the public's attention. In fake tweet Figure \ref{fake1} by resembling a terrorist's backpack to another person, he tried to introduce another person as a terrorist. In Figure \ref{fake2}, by using the image of the actual news next to another news about a missing person, the missing person was introduced as a terrorist.

As a means to deal with these challenges, we propose an end-to-end framework referred to as Fake News Revealer (FNR): A similarity and transformer-based approach to detect multi-modal fake news in social media. In this approach, we featurize text using BERT, a language representation model, which stands for Bidirectional Encoder Representations from Transformers \cite{bert}. In addition, we use the ViT, a vision transformer model that makes use of patches of images as tokens in an NLP application and provides the sequence of linear embeddings of these patches as inputs to the transformer \cite{vit}. After these two embedding modules, we use two projection modules that project extracted features into a similar-sized array and tune weights based on the task and dataset. To optimize the model, we have used two loss functions: contrastive loss, a supervised loss between image and text, and classification loss, a cross-entropy loss between the predicted and actual label of each news item.

Our main contributions are:
\begin{itemize}
	\item  
Utilizing the transformer models for both image and text, which yields better results than other text or image classification models in fake news detection.  
\item  
Using contrastive loss between the image and text to explore the relations between them for each news item.
\item 
Outperforming the state-of-the-art multi-modal fake news detection methods on two public social media datasets.

\end{itemize}
 

\section{RELATED WORKS}

With the expansion of using social networks, automatic detection of fake news has become essential. The intentional nature of fake news and its adverse effects and implications have encouraged more researchers to focus on this issue.

In this section, we categorize the related works based on the modes of their inputs. Then, we focus on the recent multi-modals fake news methods.

\subsection{Single Modality}
In early works, only one mode of data was used to detect fake news, with textual data receiving the most attention due to its prevalence in the news. Linguistic features were utilized in \cite{fake1} to validate news on Twitter, and structural and cognitive features were extracted to detect fake news on social networks in \cite{prominent}. These methods, particularly those that utilize linguistic features, are topic-specific and cannot be generalized to all topics. Additionally, the methods in \cite{fake1}, and \cite{prominent}, do not extract features automatically, resulting in insufficient and out of proportion solutions. The authors in \cite{just_image_1} and \cite{just_image_2}, implemented machine learning models on social media images to detect fake news. \\ Using recurrent deep networks, the authors in \cite{weibo}, pioneered the use of deep models for fake news detection, demonstrating improved results.

\subsection{Multiple Modality}
The authors in \cite{vqa} employed a visual system to answer questions via deep networks for fake news detection using multi-modal data. Alternatively, \cite{mitigation} used subtitle texts in addition to images for detection of fake news.

\begin{table}[!t]
\renewcommand{\arraystretch}{1.75}
\caption{Related work comparison}
\label{comparison}
\centering
\begin{tabular}{|>{\centering\arraybackslash}m{1cm}|>{\centering\arraybackslash}m{1.5cm}|>{\centering\arraybackslash}m{1.5cm}|>{\centering\arraybackslash}m{1.7cm}|}\hline
Method           & Text Encoder                   & Image Encoder           & Fusion Type                       \\ \hline
EANN\cite{eann}             & Text-CNN                       & VGG19                  & Concat                            \\ \hline
MVAE\cite{mvae}             & BiLSTM                         & VGG19                  & auto-encoder                       \\ \hline
SpotFake\cite{spotfake}         & BERT                           & VGG19                  & Concat                            \\ \hline
CARMN\cite{crossy}            & Word level sentence embeddings & VGG19                  & Concat + Crossmodal attention     \\ \hline
AMFB\cite{amfb}             & Attention based BiLSTM         & Attention based CNN - RNN & Element wise multiplication       \\ \hline
FNR (ours) & BERT                           & Visual transformer (ViT)     & Concat + Similarity  \\ \hline
\end{tabular}
\end{table}

EANN \cite{eann} is an adversarial neural network that uses image and text in the news. It tries to solve the independent identification of the news events challenge by reducing the event's impact, and the news occurrence with an adversarial mechanism. The goal is to generalize the solution to unexpected events. This model uses the pre-trained VGG19 network models for the image, and a deep canonization network for textual properties. MVAE \cite{mvae} presents a variational auto-encoder and an encoder-decoder network to detect fake news using the learned hidden vectors. A deep BiLSTM network is used to extract textual features, and a pre-trained model is used to extract image features \cite{eann}.

SpotFake \cite{spotfake} detects fake news by embedding text and images in vectors and then fusing these vectors. CARMN \cite{crossy} utilizes a cross model attention mechanism to calculate the relationship between image and text. Then, using a self-attention mechanism it obtains the feature vectors and determine fake news using a concatenation of these feature vectors. Finally, AMFB \cite{amfb} uses attention-based BiLSTM to capture textual features and attention-based CNN-RNN blocks to capture visual features. It then uses a multi-layer perceptron to classify the calculated features.

Our work focuses on fake news detection for social media by proposing a transformer and similarity-based method. Compared with the existing multi-modal fake news detection methods for the general scenario, our approach extracts contextual features from images and considers more interactions between the multi-modal data. Table \ref{comparison} provides a comparison between the previous state of the art and the proposed method.


\section{PROPOSED METHOD}

\begin{figure*}[!t]
\centerline{
		\includegraphics[width=6in]{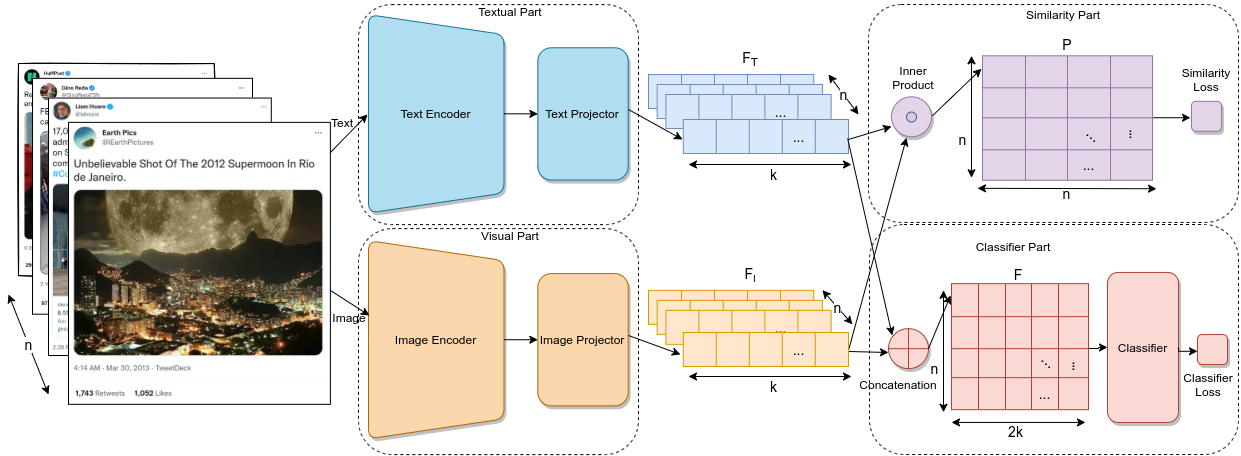}%
	}
\caption{The proposed Fake News Revealer (FNR) architecture}
\label{proposed_method}
\end{figure*}

The Fake News Revealer (FDR) architecture is shown in Figure \ref{proposed_method}. The structure of our approach is composed of four parts: 1) text, 2) image, 3) similarity, and 4) classifier. These modules are then combined to provide a label for each news. 
BERT and VGG19 are used as pre-trained models for extracting text and image feature vectors, respectively. These feature vectors are then combined and classified by a fully connected classification layer. 

We also utilize a projector module to resize the input vector for both text and image structures by applying a linear function. The projector's weights, unlike frozen encoders, are learned end-to-end during training and make the representation vector of image and text variables. Refer to Figure \ref{projection} for more details.

The input to our algorithm is $n$ news ($N$) each consists of text, image, and label (fake or real). Let $N_i$ represent news $i$, and $T_i$ , $I_i$ and $L_i$ be its text, image and label, respectively. Then:

\begin{equation}
N_i=(T_i, I_i, L_i)
\end{equation}

Our goal is to predict $L_i$ using $T_i$ and $I_i$. We process the news in a batch mode through the deep model, so in each run, the number of the input data is $b$.

\subsection{Textual Part} 

The purpose of this module is to featurize text and embed it into a vector. This part consists of two main sub-modules; the first sub-module is an encoder that extracts representative features extracted from a pre-trained model. The second component is a projector.

BERT \cite{bert} has recently been proven to be the most advanced language representation model for a variety of NLP tasks by pretraining on large corpora. During pretraining, BERT can handle short texts like tweets due to their predefined maximum length. Accordingly, the proposed model uses the BERT's textual feature extraction capabilities.

This part takes $T$ with size $(b, t)$ ($t$ is the maximum text length size). After applying BERT, we take a vector $B$ with size $(b, 768)$, where 768 is the size of the last hidden layer of BERT, and apply the projector to obtain vector $F_T$ with size $(b, k)$:

\begin{equation}
    F_T = w_2\times(gelu(w_1\times B+b_1)) + (w_1\times B+b_1) + b_2 
\end{equation}

where $w_1$, $w_2$, $b_1$ and $b_2$ are weights and biases of linear layers inside the text projector.

\subsection{Visual Part} 

This module also consists of two sub-modules. The first sub-module is an image encoder that encodes images into a representation vector. We used the ViT Vision Transformer \cite{vit} as encoder, and a transformer encoder model (BERT-like) is pre-trained on an extensive collection of images in a supervised fashion. The model is fed with fixed-size patches (resolution 16x16) linearly embedded in images. Using this pre-trained model, we are able to optimize it to our model as the second sub-module. The second sub-module is a projector, as described before.

Subsequently, this module takes $I$ with parameters $(b, width, height, depth)$ representing width, height and depth of images, and ViT project it into the vector space $V$ with size $(b, 768)$, where 768 is the size of last hidden vector of ViT model. After applying the projector, we obtain $F_i$ with size $(b, k)$ with the same length as $F_t$ that is obtain from the textual part. The projector works according to the following formula: 
\begin{equation}
    F_I = w_4\times(gelu(w_3\times V+b_3)) + (w_3\times V+b_3) + b_4
\end{equation}

where $w_3$, $w_4$, $b_3$ and $b_4$ are weights and biases of linear layers inside the image projector. 

\subsection{Similarity Part} 
In this module, we calculate the similarity of texts and images via a supervised contrastive loss \cite{clip}. We have image and text features that are matrices with size $(b, k)$, and to determine if they are similar, their inner products are often calculated.  The similarities between text and image of b tweets are represented in the $(b, b)$ matrix that we call the predicted matrix ($P$):

\begin{equation}
P = F_T{F_I}^T
\end{equation}

This loss function considers an image and a text to be the most similar to itself. Thus, we consider the expected matrix as the average similarity of the text to text and the image to image according to the following formula \cite{keras}: 
 
 \begin{equation}
 E = softmax(\dfrac{F_T{F_T}^T + F_I{F_I}^T}{2})
 \end{equation}
 
 After calculating the expected matrix ($E$), we use cross-entropy to find the actual loss. The contrastive loss is the average of the text similarity loss $l_T$, and the image similarity loss $l_I$ \cite{clip}:
 
 \begin{equation}
 l_T = -(E\times log(P)+(1-E)\times log(1-P))
 \end{equation}
  \begin{equation}
  l_I = -(E^T\times log(P^T)+(1-E^T)\times log(1-P^T))
 \end{equation}
 \begin{equation}
 l_s = \dfrac{l_T+l_I}{2}
 \end{equation}
 
\begin{figure}
\centerline{
\subfloat[Projector]
	{
	\includegraphics[width=1.1in,height=2.6in]{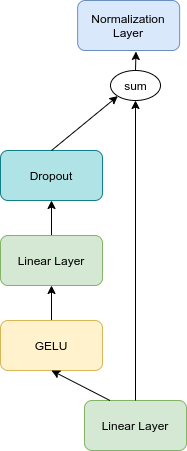}%
	\label{projection}
	}	
\hfil

\subfloat[Classifier]
	{
	\includegraphics[width=0.6in,height=2.6in]{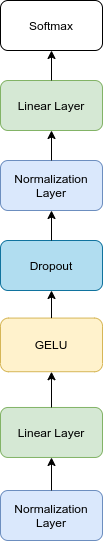}%
	\label{classification}
	}
}
\caption{The sub-modules of projector}
\label{sub-modules}
\end{figure}

\subsection{Classifier Part} 

In this module, the text and image feature vectors are concatenated to obtain the desired news representation:

\begin{equation}
F_c = concat(F_T, F_I)
\end{equation}

This news representation is then passed through two linear layers for fake news classification. After this linear mapping, we consider a vector of size $(b, 2)$ with two classes. According to Figure \ref{classification} and assuming $w_6$, $w_5$, $b_6$ and $b_5$ are weights and biases of linear layers inside the classifier:

\begin{equation}
    F = softmax(w_6\times(gelu(w_5\times F_c+b_5)) + b_6)
\end{equation}

After passing the vector through the softmax function, we optimize the model by calculating the predicted and actual labels' cross-entropy ($\alpha$ is the balancing factor for the weighted loss function):

\begin{equation}
l_c = -(\alpha L\times log(F) + (1-L)\times log(1-F))
\end{equation}

The loss for the whole network is obtained from the two losses with $\lambda$ as a tradeoff parameter:

\begin{equation}
l = l_c +  \lambda \cdot l_s.
\end{equation}

\section{EXPRIMENTS}
This section presents the implementation issues and experimental results of applying FNR to real datasets and its comparison with the state of the art methods. 

\subsection{Datasets}
We used two publicly available datasets gathered from social media as described below.

\subsubsection{Twitter} This dataset was introduced in \cite{Twitter} for automatic verification of multimedia tasks to distinguish fake/real news on Twitter\footnote{https://twitter.com/}. Data consists of a development set and a test set, each with its own events. The rows of data contain text, images or video, and additional information about the user's profile. The tweets in this dataset are mainly in English (other languages were translated to English before running the experiments).

\subsubsection{Weibo} 
This dataset \cite{weibo}  was collected from Weibo\footnote{https://weibo.com/} social media from 2012 to 2016 and is written in the Chinese language. Each line of data in this dataset also contains text, user information, and an image. The app's authentication system has tagged texts. This database is divided by \cite{eann} into two sets of test data and the train, as the news events of each set are different and. We utilized the same test and train sets.

Only the data with text and images are used in both datasets, and only the first image was utilized. Table \ref{datasets} lists the number of train/test data and fake/real data that have been used in our experiments. 

\begin{table}[]
\renewcommand{\arraystretch}{1.3}
\caption{The distribution of datasets}
\label{datasets}
\centering
\begin{tabular}{|cc|c|c|}
\hline
\multicolumn{1}{|c|}{}                       &      & Twitter & Weibo \\ \hline
\multicolumn{1}{|c|}{\multirow{2}{*}{Train}} & Fake & 6649    & 3748  \\ \cline{2-4} 
\multicolumn{1}{|c|}{}                       & Real & 4599    & 3758  \\ \hline
\multicolumn{1}{|c|}{\multirow{2}{*}{Test}}  & Fake & 545     & 999   \\ \cline{2-4} 
\multicolumn{1}{|c|}{}                       & Real & 444     & 995   \\ \hline
\multicolumn{2}{|c|}{All data}                      & 12237   & 9500  \\ \hline
\end{tabular}
\end{table}

\subsection{Implementation Details}
The Pytorch framework is used to build our architecture with Python 3.6. We optimized the learning rate using AdamW and calculated different learning rates and weight decays for each part of our architecture to make model converge faster. We have used the Optuna\footnote{https://optuna.org/} library to find the best parameters. The best parameters are as follows: classification learning rate of 0.005, classification weight decay of 0.07, projection vector size ($k$) of 64, dropout of 0.3, $\lambda$ was set to 1, and $\alpha$ was the ratio of class data with a large number versus class data with a smaller number.

We set the batch size to 256, our epoch number was set to 100, and the maximum text length ($t$) for Twitter was 32 words and for Weibo was 200 characters. All input images resized to height=224, width=224 and depth=3. We used a learning rate scheduler and an early stopping checkpoint to avoid overfitting. For the encoder models, the Hugging Face\footnote{https://huggingface.co/} library was used. The implementation is available in our repository\footnote{http://git.dml.ir/fghorbanpoor/FakeNewsRevealer}.

Because the input text is raw text gathered from social media, it is non-standard and noisy, and it needs to be cleaned up using normalization techniques. We performed preprocessing which includes converting abbreviations to complete the forms, removing unnecessary punctuations, and deleting non-standard characters. Texts in various languages were also translated to match the dataset's language. Our text is subsequently tokenized and made ready to be encoded after the preprocessing step. The images were also preprocessed by deleting low-quality images, resizing them to (224x224), and converting them into appropriate input for the encoder.

\subsection{Baselines}
The following are some benchmark algorithms and models that we chose for a comparative study.
\subsubsection{Single Modality}
We just considered news text and examined some algorithms to achieve a better feature vector in the textual module. These parameters were tuned by Optuna, and the best results are reported.
Our text is subsequently tokenized and ready to be encoded after the preprocessing step.

For the visual module, we used images of news and tuned the parameters with Optuna.  The CNN architecture is three parallel convolutional layers with different filter sizes were utilized in this step.

\subsubsection{Multi-Modality}
For multi-modality, we have chosen recent works that are state of the art. These works are listed iin Table \ref{comparison}.

We tested two versions of the proposed model in this section; one without considering the similarity measurements (FNR-WS), and the other with considering similarity measurements (FNR-S). 

\subsection{Results}

To compare the proposed method in solving multi-modal fake news detection with the previous works, we considered the following criteria for evaluation; accuracy, recall, precision, and F1-score. When dealing with classification problems, these criteria are always taken into consideration. 

We are trying to solve a binary classification with balanced data, since the number of fake news and real news are almost equal. We can also consider the AUC criterion and the receiver operating characteristic (ROC) curve for classification evaluation. We included their measurement in the comparison table and plotted the curves for three previous works for comparison purposes.

\subsubsection{Ablation Study}

 \begin{figure}
	\includegraphics[width=3in, height=2in]{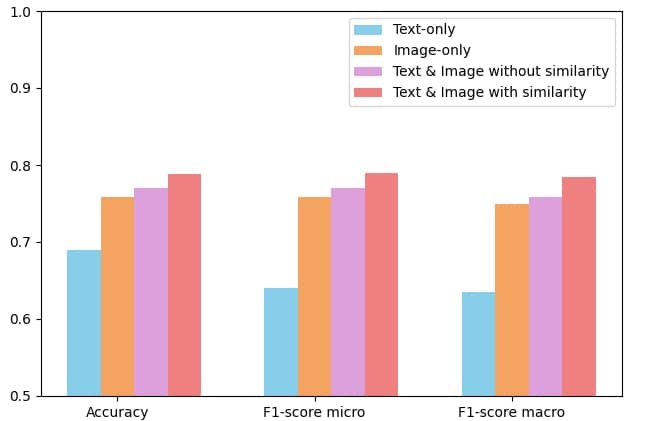}
\caption{Ablation study on Twitter dataset}
\label{twitter_ablation}
\end{figure}

 \begin{figure}
	\includegraphics[width=3in, height=2in]{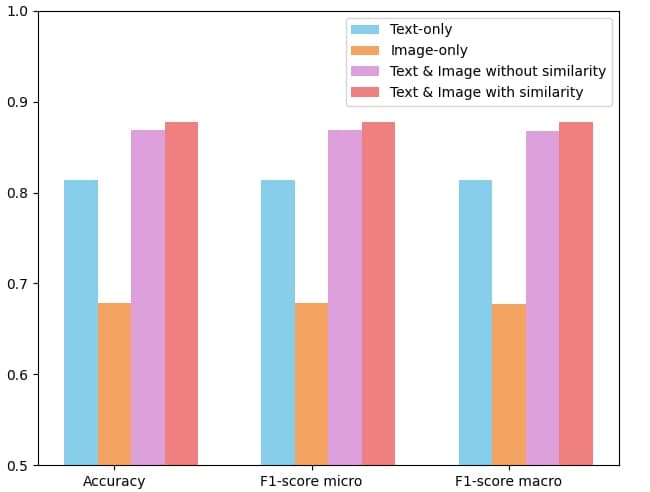}
\caption{Ablation study on Weibo dataset}
\label{weibo_ablation}
\end{figure}

An ablation study was performed to explore which data mode is more beneficial or why a multi-modal approach should be used and why we should consider similarity measures. First, we used and tested each mode separately, and then we fused the modes and investigated the effectiveness of a multi-modal approach for fake news detection. Finally, we added the contrastive loss measurement to investigate its effectiveness in enhancing the results.

As demonstrated by Figure \ref{twitter_ablation} on Twitter, because the language of the tweets is brief, imprecise, and filthy, it is less accurate on its own, and the image mode performed better. However, the outcome improves by merging these two modalities, demonstrating that these two modes cover each other's shortcomings. Adding the similarity part, which considers the relationship between the text and the image, results in more improvements.

As illustrated in Figure \ref{weibo_ablation}, the images on Weibo are not very expressive, and the actual news images are almost certainly being exploited for false news, which does not help to detect fake news on its own. Nonetheless, the text mode outperforms the visual mode. However, when these two modalities are included concurrently, the model's performance significantly improves, as these two type of modalities complement each other. When the relationship between text and image is considered, the accuracy also increases.

\subsubsection{Performance Comparison}
\begin{figure}
	\includegraphics[width=2.8in, height=1.8in]{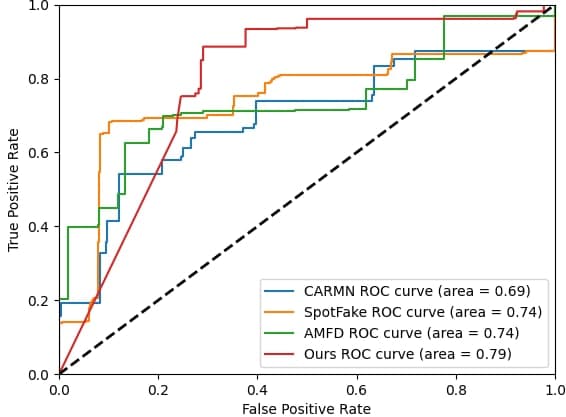}
\caption{ROC curve on Twitter dataset}
\label{twitter_auc}
\end{figure}

Using the mentioned criteria, the following comparison table for the Twitter dataset is presented in Table \ref{twitter_comparison} and for the Weibo dataset in Table \ref{weibo_comparison} and Figure \ref{twitter_auc} compares the ROC curves of three relevant studies and our work on the Twitter dataset, and Figure 7 compares them on the Weibo dataset.

\begin{table*}[!t]
\renewcommand{\arraystretch}{1.3}
\caption{Comparison for the Twitter dataset}
\label{twitter_comparison}
\centering
\begin{tabular}{|>{\centering\arraybackslash}m{1.2cm}|>{\centering\arraybackslash}m{1.3cm}|c|c|>{\centering\arraybackslash}m{1cm}|ccc|ccc|}
\hline
                            &                                              &                &           &     & \multicolumn{3}{c|}{Fake}                                                               & \multicolumn{3}{c|}{Real}                                                               \\ \hline
                            & Model Name                                   & Accuracy       & AUC  & F1-score Micro          & \multicolumn{1}{c|}{Precision}     & \multicolumn{1}{c|}{Recall}        & F1-score      & \multicolumn{1}{c|}{Precision}     & \multicolumn{1}{c|}{Recall}        & F1-score      \\ \hline \hline
\multirow{4}{*}{Text}       & Logistic Regression                          & 0.62          & 0.46   &0.62       & \multicolumn{1}{c|}{0.69}          & \multicolumn{1}{c|}{0.55}          & 0.61          & \multicolumn{1}{c|}{0.56}          & \multicolumn{1}{c|}{0.70}          & 0.62          \\ \cline{2-11} 
                            & SVM                                          & 0.61          & 0.46   &0.61       & \multicolumn{1}{c|}{0.68}          & \multicolumn{1}{c|}{0.55}          & 0.61          & \multicolumn{1}{c|}{0.55}          & \multicolumn{1}{c|}{0.68}          & 0.62          \\ \cline{2-11} 
                            & BiLSTM                                       & 0.61          & 0.59  &0.60        & \multicolumn{1}{c|}{0.62}          & \multicolumn{1}{c|}{0.73}          & 0.67          & \multicolumn{1}{c|}{0.58}          & \multicolumn{1}{c|}{0.45}          & 0.51          \\ \cline{2-11} 
                            & BERT                                         & 0.69          & 0.69      & 0.64    & \multicolumn{1}{c|}{0.67}          & \multicolumn{1}{c|}{0.68}          & 0.68          & \multicolumn{1}{c|}{0.60}          & \multicolumn{1}{c|}{0.59}          & 0.59          \\ \hline \hline
\multirow{3}{*}{Image}      & CNN                                          &  0.62         & 0.46     & 0.62     & \multicolumn{1}{c|}{0.69}          & \multicolumn{1}{c|}{0.55}          & 0.61              & \multicolumn{1}{c|}{0.56}          & \multicolumn{1}{c|}{0.55}          &    0.61         \\ \cline{2-11} 
                            & VGG19                                       & 0.68          & 0.46   &0.68       & \multicolumn{1}{c|}{0.74}          & \multicolumn{1}{c|}{0.64}          & 0.69          & \multicolumn{1}{c|}{0.62}          & \multicolumn{1}{c|}{0.72}          & 0.67          \\ \cline{2-11} 
                            & ViT                                          & 0.72 & 0.71   & 0.76       & \multicolumn{1}{c|}{0.74}          & \multicolumn{1}{c|}{0.86}          & 0.80          & \multicolumn{1}{c|}{0.79}          & \multicolumn{1}{c|}{0.63}          & 0.70          \\ \hline \hline
\multirow{7}{*}{multi-modal} & EANN                                         & 0.69          &   0.72    & 0.69        & \multicolumn{1}{c|}{0.75}          & \multicolumn{1}{c|}{0.58}          & 0.65          & \multicolumn{1}{c|}{0.62}          & \multicolumn{1}{c|}{0.76}          & 0.69          \\ \cline{2-11} 
                            & MVAE                                         & 0.67          &  0.66   & 0.67          & \multicolumn{1}{c|}{0.70}          & \multicolumn{1}{c|}{0.69}          & 0.69          & \multicolumn{1}{c|}{0.63}          & \multicolumn{1}{c|}{0.64}          & 0.63          \\ \cline{2-11} 
                            & SpotFake                                     & 0.77          & 0.74    & 0.76      & \multicolumn{1}{c|}{0.72}          & \multicolumn{1}{c|}{\textbf{0.92}} & 0.81          & \multicolumn{1}{c|}{\textbf{0.85}} & \multicolumn{1}{c|}{0.56}          & 0.68          \\ \cline{2-11} 
                            & CARMN                                        & 0.73          & 0.69    &0.73      & \multicolumn{1}{c|}{0.70}          & \multicolumn{1}{c|}{0.88}          & 0.78          & \multicolumn{1}{c|}{0.78}          & \multicolumn{1}{c|}{0.54}          & 0.64          \\ \cline{2-11} 
                            & AMFD                                         & 0.75          & 0.74    &0.75      & \multicolumn{1}{c|}{0.76}          & \multicolumn{1}{c|}{0.79}          & 0.78          & \multicolumn{1}{c|}{0.73}          & \multicolumn{1}{c|}{0.70}          & 0.71          \\ \cline{2-11} 
                            & FNR-WS & 0.77          & 0.76   & 0.77       & \multicolumn{1}{c|}{0.74}          & \multicolumn{1}{c|}{0.90}          & 0.81          & \multicolumn{1}{c|}{0.83}          & \multicolumn{1}{c|}{0.62}          & 0.71          \\ \cline{2-11} 
                            & FNR-S    & \textbf{0.79} & \textbf{0.79} & \textbf{0.79} & \multicolumn{1}{c|}{\textbf{0.78}} & \multicolumn{1}{c|}{0.85}          & \textbf{0.82} & \multicolumn{1}{c|}{0.79}          & \multicolumn{1}{c|}{\textbf{0.71}} & \textbf{0.75} \\ \hline
\end{tabular}
\end{table*}

\begin{table*}[!t]
\renewcommand{\arraystretch}{1.3}
\caption{Comparison for the Weibo dataset}
\label{weibo_comparison}
\centering
\begin{tabular}{|>{\centering\arraybackslash}m{1.2cm}|>{\centering\arraybackslash}m{1.3cm}|c|c|>{\centering\arraybackslash}m{1cm}|ccc|ccc|}
\hline
                               &                           &                &        &        & \multicolumn{3}{c|}{Fake}                                                               & \multicolumn{3}{c|}{Real}                                                               \\ \hline
                               & Model Name                & Accuracy       & AUC   & F1-score Micro         & \multicolumn{1}{c|}{Precision}     & \multicolumn{1}{c|}{Recall}        & F1-score      & \multicolumn{1}{c|}{Precision}     & \multicolumn{1}{c|}{Recall}        & F1-score      \\ \hline \hline
\multirow{4}{*}{Text}          & Logistic Regression       & 0.71          & 0.67    & 0.71      & \multicolumn{1}{c|}{0.71}          & \multicolumn{1}{c|}{0.80}          & 0.75          & \multicolumn{1}{c|}{0.71}          & \multicolumn{1}{c|}{0.59}          & 0.65          \\ \cline{2-11} 
                               & SVM                       & 0.70          & 0.70    &0.70      & \multicolumn{1}{c|}{0.72}          & \multicolumn{1}{c|}{0.74}          & 0.73          & \multicolumn{1}{c|}{0.67}          & \multicolumn{1}{c|}{0.65}          & 0.66          \\ \cline{2-11} 
                               & BiLSTM                    & 0.66          & 0.44     &0.66     & \multicolumn{1}{c|}{0.62}          & \multicolumn{1}{c|}{0.78}          &  0.69         & \multicolumn{1}{c|}{0.73}          & \multicolumn{1}{c|}{0.55}          & 0.63              \\ \cline{2-11} 
                               & BERT                      & 0.81          & 0.83   & 0.81       & \multicolumn{1}{c|}{0.81}          & \multicolumn{1}{c|}{0.81}          & 0.81          & \multicolumn{1}{c|}{0.81}          & \multicolumn{1}{c|}{0.82}          & 0.81          \\ \hline \hline
\multirow{3}{*}{Image}         & CNN                       & 0.52          & 0.39    & 0.50      & \multicolumn{1}{c|}{0.79}          & \multicolumn{1}{c|}{0.24}          & 0.38          & \multicolumn{1}{c|}{0.58}          & \multicolumn{1}{c|}{0.87}          & 0.63          \\ \cline{2-11} 
                               & VGG19                    & 0.60          & 0.47    & 0.60      & \multicolumn{1}{c|}{0.60}          & \multicolumn{1}{c|}{0.61}          & 0.60          & \multicolumn{1}{c|}{0.60}          & \multicolumn{1}{c|}{0.59}          & 0.59          \\ \cline{2-11} 
                               & ViT                       & 0.68          & 0.72    & 0.68      & \multicolumn{1}{c|}{0.67}          & \multicolumn{1}{c|}{0.69}          & 0.68          & \multicolumn{1}{c|}{0.68}          & \multicolumn{1}{c|}{0.68}          & 0.67          \\ \hline \hline
\multirow{7}{*}{multi-modal} & EANN                      & 0.81          &  0.86   & 0.81        & \multicolumn{1}{c|}{0.89}          & \multicolumn{1}{c|}{0.66}          & 0.76          & \multicolumn{1}{c|}{0.77}          & \multicolumn{1}{c|}{0.93}          & 0.85          \\ \cline{2-11} 
                               & MVAE                      & 0.79          &  0.79   & 0.79      & \multicolumn{1}{c|}{0.89}          & \multicolumn{1}{c|}{0.65}          & 0.75          & \multicolumn{1}{c|}{0.74}          & \multicolumn{1}{c|}{0.93}          & 0.82          \\ \cline{2-11} 
                               & SpotFake                  & 0.86          & 0.90   & 0.86       & \multicolumn{1}{c|}{0.87}          & \multicolumn{1}{c|}{0.92}          & 0.90          & \multicolumn{1}{c|}{0.81}          & \multicolumn{1}{c|}{0.70}          & 0.75          \\ \cline{2-11} 
                               & CARMN                     & 0.84          & 0.90   & 0.85       & \multicolumn{1}{c|}{0.86}          & \multicolumn{1}{c|}{\textbf{0.93}} & \textbf{0.89} & \multicolumn{1}{c|}{0.81}          & \multicolumn{1}{c|}{0.66}          & 0.73          \\ \cline{2-11} 
                               & AMFD                      & 0.83          & 0.89  & 0.83        & \multicolumn{1}{c|}{0.86}          & \multicolumn{1}{c|}{0.90}          & 0.88          & \multicolumn{1}{c|}{0.75}          & \multicolumn{1}{c|}{0.68}          & 0.71          \\ \cline{2-11} 
                               & FNR-WS & 0.87          & 0.90     & 0.87     & \multicolumn{1}{c|}{0.89}          & \multicolumn{1}{c|}{0.85}          & 0.87          & \multicolumn{1}{c|}{0.85}          & \multicolumn{1}{c|}{0.89}          & 0.87          \\ \cline{2-11} 
                               & FNR-S   & \textbf{0.88} & \textbf{0.94} & \textbf{0.88} &\multicolumn{1}{c|}{\textbf{0.89}} & \multicolumn{1}{c|}{0.87}          & 0.88          & \multicolumn{1}{c|}{\textbf{0.87}} & \multicolumn{1}{c|}{\textbf{0.89}} & \textbf{0.88} \\ \hline
\end{tabular}
\end{table*}

 \begin{figure}
	\includegraphics[width=2.8in, height=1.8in]{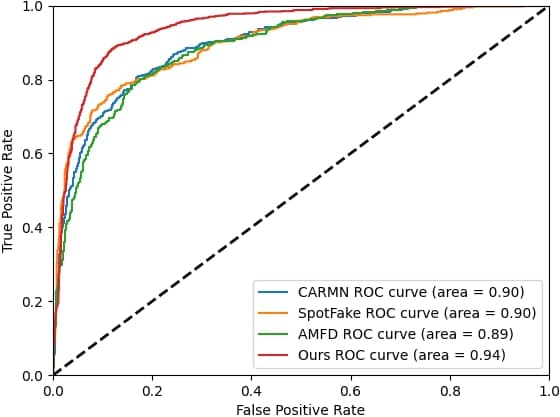}
\caption{ROC curve on Weibo dataset}
\label{weibo_auc}
\end{figure}

Based on the results in these tables, BERT has performed better than the other text algorithms. It has properly extracted conceptual and lexical features, and has a more accurate approach in detecting fake news with text only. 

Moreover, ViT outperforms other fake news detection algorithms that rely solely on images, and with a transformer-based network, it extracts valuable and efficient features.

According to the results, some current methods in multi-modal fake news detection perform more favorably compare to single mode methods, indicating that the combination of data modes and the extraction of their intermodal properties are beneficial. Spotfake outperformed the competition because it pulls more useful characteristics from text and images. In addition, both CARMN and AMFD have presented a more performant combination of attention-based mechanisms.

The Fake News Revealer (FNR) framework consistently outperforms the competition on a variety of performance metrics. Each modality can retain its distinct traits with our methodology while seamlessly combining similarity and complementary information from the other modalities.


\section{CONCLUSION AND FUTURE WORKS}
A new multi-modal framework for detecting fake news employing text and image transformers, as well as similarity measures with a contrastive loss function, is presented in this work. To extract the textual information, we employed BERT, which is a transformer-based encoder. The visual characteristics were retrieved using ViT, a BERT-like transformer that was explicitly created for image processing. The connection between an image and a sentence is extracted using a contrastive loss and then fed into a two-layer linear classifier to assess if the news is real or fake. We tested the proposed method on two freely available multi-modal datasets. Extensive tests have been carried out to compare to existing models. The suggested framework (FNR) outperforms the state of the art methods in detecting fake news. Since a large amount of news is being published with different modes, one can use other data modes such as video, audio, and news graphs. 



\bibliographystyle{IEEEtran}
\bibliography{references}

\begin{thebibliography}{10}
\providecommand{\url}[1]{#1}
\csname url@samestyle\endcsname
\providecommand{\newblock}{\relax}
\providecommand{\bibinfo}[2]{#2}
\providecommand{\BIBentrySTDinterwordspacing}{\spaceskip=0pt\relax}
\providecommand{\BIBentryALTinterwordstretchfactor}{4}
\providecommand{\BIBentryALTinterwordspacing}{\spaceskip=\fontdimen2\font plus
\BIBentryALTinterwordstretchfactor\fontdimen3\font minus
  \fontdimen4\font\relax}
\providecommand{\BIBforeignlanguage}[2]{{%
\expandafter\ifx\csname l@#1\endcsname\relax
\typeout{** WARNING: IEEEtran.bst: No hyphenation pattern has been}%
\typeout{** loaded for the language `#1'. Using the pattern for}%
\typeout{** the default language instead.}%
\else
\language=\csname l@#1\endcsname
\fi
#2}}
\providecommand{\BIBdecl}{\relax}
\BIBdecl

\bibitem{news1}
\BIBentryALTinterwordspacing
D.~Gross, ``Survey: More americans get news from internet than newspapers or
  radio,'' accessed:2020-01-16. [Online]. Available:
  \url{http://www.cnn.com/2010/TECH/03/01/social.network.news/index.html}
\BIBentrySTDinterwordspacing

\bibitem{fake1}
K.~Shu, A.~Sliva, S.~Wang, J.~Tang, and H.~Liu, ``Fake news detection on social
  media: A data mining perspective,'' \emph{SIGKDD Explor. Newsl.}, vol.~19,
  no.~1, p. 22–36, Sep 2017.

\bibitem{image1}
E.~J. Newman, M.~Garry, D.~M. Bernstein, J.~Kantner, and D.~S. Lindsay,
  ``Nonprobative photographs (or words) inflate truthiness,'' \emph{Psychonomic
  Bulletin {\&} Review}, vol.~19, no.~5, pp. 969--974, Oct 2012.

\bibitem{image2}
E.~Fenn, N.~Ramsay, J.~Kantner, K.~Pezdek, and E.~Abed, ``Nonprobative photos
  increase truth, like, and share judgments in a simulated social media
  environment,'' \emph{JARMAC}, vol.~8, no.~2, pp. 131--138, 2019.

\bibitem{bert}
J.~Devlin, M.~Chang, K.~Lee, and K.~Toutanova, ``{BERT:} pre-training of deep
  bidirectional transformers for language understanding,'' in \emph{Proc. North
  Amer. Ch. Assoc. Comput. Linguistics: Human Lang. Technol}, 2019, pp.
  4171--4186.

\bibitem{vit}
\BIBentryALTinterwordspacing
A.~Dosovitskiy, L.~Beyer, A.~Kolesnikov, D.~Weissenborn, X.~Zhai,
  T.~Unterthiner, M.~Dehghani, M.~Minderer, G.~Heigold, S.~Gelly, J.~Uszkoreit,
  and N.~Houlsby, ``An image is worth 16x16 words: Transformers for image
  recognition at scale,'' 2021. [Online]. Available:
  \url{https://arxiv.org/abs/2010.11929}
\BIBentrySTDinterwordspacing

\bibitem{prominent}
S.~Kwon, M.~Cha, K.~Jung, W.~Chen, and Y.~Wang, ``Prominent features of rumor
  propagation in online social media,'' in \emph{Proc. 13th IEEE Conf. Data
  Min.}, 2013, pp. 1103--1108.

\bibitem{just_image_1}
M.~Gupta, P.~Zhao, and J.~Han, ``Evaluating event credibility on twitter,'' in
  \emph{Proc. SIAM Conf. Data Min.}, 2012, pp. 153--164.

\bibitem{just_image_2}
D.~Tian, ``A review on image feature extraction and representation
  techniques,'' \emph{Int. J.}, vol.~8, pp. 385--395, Jan 2013.

\bibitem{weibo}
Z.~Jin, J.~Cao, H.~Guo, Y.~Zhang, and J.~Luo, ``Multimodal fusion with
  recurrent neural networks for rumor detection on microblogs,'' in \emph{Proc.
  25th ACM Conf. Multimedia}, 2017, pp. 795--816.

\bibitem{vqa}
S.~Antol, A.~Agrawal, J.~Lu, M.~Mitchell, D.~Batra, C.~L. Zitnick, and
  D.~Parikh, ``Vqa: Visual question answering,'' in \emph{Proc. IEEE Conf.
  Comput. Vis.}, 2015, pp. 2425--2433.

\bibitem{mitigation}
M.~Farajtabar, J.~Yang, X.~Ye, H.~Xu, R.~Trivedi, E.~Khalil, S.~Li, L.~Song,
  and H.~Zha, ``Fake news mitigation via point process based intervention,'' in
  \emph{Proc. 34th Conf. Mach. Learn. Res.}, 2017, p. 1097–1106.

\bibitem{eann}
Y.~Wang, F.~Ma, Z.~Jin, Y.~Yuan, G.~Xun, K.~Jha, L.~Su, and J.~Gao, ``Eann:
  Event adversarial neural networks for multi-modal fake news detection,'' in
  \emph{Proc. 24th ACM SIGKDD Conf. Data Min. Knowl. Discov.}, 2018, p.
  849–857.

\bibitem{mvae}
D.~Khattar, J.~S. Goud, M.~Gupta, and V.~Varma, ``Mvae: Multimodal variational
  autoencoder for fake news detection,'' in \emph{Proc. 19th WWW Conf.}, 2019,
  p. 2915–2921.

\bibitem{spotfake}
S.~Singhal, R.~R. Shah, T.~Chakraborty, P.~Kumaraguru, and S.~Satoh,
  ``Spotfake: A multi-modal framework for fake news detection,'' in \emph{Proc.
  IEEE 5th Conf. Multimedia BigMM}, 2019, pp. 39--47.

\bibitem{crossy}
C.~Song, N.~Ning, Y.~Zhang, and B.~Wu, ``A multimodal fake news detection model
  based on crossmodal attention residual and multichannel convolutional neural
  networks,'' \emph{Inf. Process. Manag.}, vol.~58, no.~1, p. 102437, 2021.

\bibitem{amfb}
R.~Kumari and A.~Ekbal, ``Amfb: Attention based multimodal factorized bilinear
  pooling for multimodal fake news detection,'' \emph{Expert Syst. Appl.}, vol.
  184, p. 115412, 2021.

\bibitem{clip}
\BIBentryALTinterwordspacing
A.~Radford, J.~W. Kim, C.~Hallacy, A.~Ramesh, G.~Goh, S.~Agarwal, G.~Sastry,
  A.~Askell, P.~Mishkin, J.~Clark, G.~Krueger, and I.~Sutskever, ``Learning
  transferable visual models from natural language supervision,'' 2021.
  [Online]. Available: \url{https://arxiv.org/abs/2103.00020}
\BIBentrySTDinterwordspacing

\bibitem{keras}
\BIBentryALTinterwordspacing
K.~Salama, ``Keras documentation: Natural language image search with a dual
  encoder,'' accessed:2021-11-08. [Online]. Available:
  \url{https://keras.io/examples/nlp/nl_image_search/}
\BIBentrySTDinterwordspacing

\bibitem{Twitter}
C.~Boididou, K.~Andreadou, S.~Papadopoulos, D.~T. Dang~Nguyen, G.~Boato,
  M.~Riegler, M.~Larson, and I.~Kompatsiaris, ``Verifying multimedia use at
  mediaeval 2015,'' in \emph{MediaEval}, Sep 2015.

\end{thebibliography}

\begin{IEEEbiography}{Faeze Ghorbanpour}{\,} is currently a research fellow at the Digital Media Laboratory, Sharif University of Technology. Her research interests include computational social science and social networks. She received her B.Sc. and M.Sc. degrees in computer engineering from the Sharif University of Technology in Tehran, Iran, in 2018 and 2021, respectively.
\end{IEEEbiography}

\begin{IEEEbiography}{Maryam Ramezani}{\,}
  received her B.Sc. and M.Sc. degrees in Information Technology Engineering from the Sharif University of Technology, Tehran, Iran. She is currently a PhD candidate in the Department of Computer Engineering at the Sharif University of Technology. Her current research interests include machine learning with application in social and complex networks.
\end{IEEEbiography}

\begin{IEEEbiography}{MohammadAmin Fazli}{\,}
received his B.Sc. degree in hardware engineering and the M.Sc. and PhD degrees in software engineering from the Sharif University of Technology in 2009, 2011, and 2015. He is currently an Assistant Professor at the Sharif University of Technology and the R\&D supervisor of the Intelligent Information Center. His research interests include game theory, combinatorial optimization, computational economics, graphs and combinatorics, complex networks, and dynamical systems.
\end{IEEEbiography}

\begin{IEEEbiography}{Hamid R. Rabiee}{\,} 
 received his BS and MS degrees (with Great Distinction) in Electrical Engineering from CSULB, Long Beach, CA (1987, 1989), his EEE degree in Electrical and Computer Engineering from USC, Los Angeles, CA (1993), and his Ph.D. in Electrical and Computer Engineering from Purdue University, West Lafayette, IN, in 1996. From 1993 to 1996 he was a Member of Technical Staff at AT\&T Bell Laboratories. From 1996 to 1999 he worked as a Senior Software Engineer at Intel Corporation. He was also with PSU, OGI and OSU universities as an adjunct professor of Electrical and Computer Engineering from 1996-2000. Since September 2000, he has joined Sharif University of Technology, Tehran, Iran. He was also a visiting professor at the Imperial College of London for the 2017-2018 academic year. He is the founder of Sharif University Advanced Information and Communication Technology Research Institute (AICT), ICT Innovation Center, Advanced Technologies Incubator (SATI), Digital Media Laboratory (DML), Mobile Value Added Services Laboratory (VASL), Bioinformatics and Computational Biology Laboratory (BCB) and Cognitive Neuroengineering Research Center. He is also a consultant and member of AI in Health Expert Group at WHO. He has been the founder of many successful High-Tech start-up companies in the field of ICT as an entrepreneur. He is currently a Professor of Computer Engineering at Sharif University of Technology, and Director of AICT, DML, and VASL. He has received numerous awards and honors for his Industrial, scientific and academic contributions, and holds three patents. His research interests include statistical machine learning, Bayesian statistics, data analytics and complex networks with applications in social networks, multimedia systems, cloud and IoT privacy, bioinformatics, and brain networks.
\end{IEEEbiography}

\end{document}